\documentclass[a4paper,usenatbib,times]{mn2e}
\usepackage{natbib,graphicx,amsmath,amsfonts,amssymb,times,txfonts,color,bm}

\def\sl{S_{\mathrm{l}}}

\def\sopt{s_{\mathrm{opt}}}
\def\rhog{\rho_{\mathrm{g}}}
\def\rhod{\rho_{\mathrm{d}}}
\def\ts{t_{\mathrm{s}}}

\def\tk{t_{\mathrm{k}}}

\def\vkz{v_{\mathrm{k,0}}}

\def\sz{S_{\mathrm{0}}}

\def\etaz{\eta_{\mathrm{0}}}

\def\dst{\displaystyle}
\def\St{\mathrm{St}}

\def\rhog{\rho_{\mathrm{g}}}
\def\rhod{\rho_{\mathrm{d}}}
\def\cs{c_{\mathrm{s}}}
\def\brhog{\bar{\rho}_{\mathrm{g}}}
\def\bcs{\bar{c}_\mathrm{s}}

\def\Rz{r_{0}}
\def\csz{c_{\mathrm{s}0}}

\def\Sigmaz{\Sigma_{0}}
\def\vkz{v_{\mathrm{k}0}}

\def\Hz{H_{0}}

\def\tsz{t_{\mathrm{s}0}}

\def\md{m_{\mathrm{d}}}
\def\ts{t_{\mathrm{s}}}

\def\tvr{\tilde{v}_{r}}
\def\tvtheta{\tilde{v}_{\theta}}

\def\vkz{v_{\mathrm{k}0}}
\def\phiz{\phi_{0}}
\def\etaz{\eta_{0}}

\def\sz{S_{0}}

\def\gammacs{\gamma_{\mathrm{c,s}}}
\def\gammacm{\gamma_{\mathrm{c,m}}}

\newcommand{\ind}[2]{#1_\mathrm{#2}}

\newcommand{\ddiff}[1]{\mathrm{d} \!\!\ #1}

\title[Growing dust grains: vertical settling]{Growing dust grains in protoplanetary discs --- III. Vertical settling}

\author[G. Laibe et al.]{Guillaume Laibe$^{1,4}$\thanks{E-mail:guillaume.laibe@monash.edu}, Jean-Fran\c cois Gonzalez$^{2}$, Sarah T. Maddison$^{3}$, Elisabeth Crespe$^{2}$ \\
$^1$Monash Centre for Astrophysics (MoCA) and School of Mathematical Sciences, Monash University, Clayton, Vic 3800, Australia\\
$^2$Universit\'e de Lyon, Lyon, F-69003, France; Universit\'e Lyon 1, Villeurbanne, F-69622, France; CNRS, UMR 5574, Centre de Recherche Astrophysique de Lyon;\\
 \'Ecole normale sup\'erieure de Lyon, 46, all\'ee d'Italie, F-69364 Lyon Cedex 07, France\\
$^3$Centre for Astrophysics and Supercomputing, Swinburne University,
PO Box 218, Hawthorn, VIC 3122, Australia\\
$^4$School of Physics and Astronomy, University of Saint Andrews, North Haugh, St Andrews, Fife KY16 9SS
}

\pagerange{\pageref{firstpage}--\pageref{lastpage}} \pubyear{2012}

\begin{document}
%
%
%


\def\jnl@style{\it}
\def\aaref@jnl#1{{\jnl@style#1}}

\def\aaref@jnl#1{{\jnl@style#1}}

\def\aj{\aaref@jnl{AJ}}                   
\def\araa{\aaref@jnl{ARA\&A}}             
\def\apj{\aaref@jnl{ApJ}}                 
\def\apjl{\aaref@jnl{ApJ}}                
\def\apjs{\aaref@jnl{ApJS}}               
\def\ao{\aaref@jnl{Appl.~Opt.}}           
\def\apss{\aaref@jnl{Ap\&SS}}             
\def\aap{\aaref@jnl{A\&A}}                
\def\aapr{\aaref@jnl{A\&A~Rev.}}          
\def\aaps{\aaref@jnl{A\&AS}}              
\def\azh{\aaref@jnl{AZh}}                 
\def\baas{\aaref@jnl{BAAS}}               
\def\icarus{\aaref@jnl{icarus}} 
\def\jrasc{\aaref@jnl{JRASC}}             
\def\memras{\aaref@jnl{MmRAS}}            
\def\mnras{\aaref@jnl{MNRAS}}             
\def\pra{\aaref@jnl{Phys.~Rev.~A}}        
\def\prb{\aaref@jnl{Phys.~Rev.~B}}        
\def\prc{\aaref@jnl{Phys.~Rev.~C}}        
\def\prd{\aaref@jnl{Phys.~Rev.~D}}        
\def\pre{\aaref@jnl{Phys.~Rev.~E}}        
\def\prl{\aaref@jnl{Phys.~Rev.~Lett.}}    
\def\pasp{\aaref@jnl{PASP}}               
\def\pasj{\aaref@jnl{PASJ}}               
\def\qjras{\aaref@jnl{QJRAS}}             
\def\skytel{\aaref@jnl{S\&T}}             
\def\solphys{\aaref@jnl{Sol.~Phys.}}      
\def\sovast{\aaref@jnl{Soviet~Ast.}}      
\def\ssr{\aaref@jnl{Space~Sci.~Rev.}}     
\def\zap{\aaref@jnl{ZAp}}                 
\def\nat{\aaref@jnl{Nature}}              
\def\iaucirc{\aaref@jnl{IAU~Circ.}}       
\def\aplett{\aaref@jnl{Astrophys.~Lett.}} 
\def\apspr{\aaref@jnl{Astrophys.~Space~Phys.~Res.}}
\def\bain{\aaref@jnl{Bull.~Astron.~Inst.~Netherlands}} 
\def\fcp{\aaref@jnl{Fund.~Cosmic~Phys.}}  
\def\gca{\aaref@jnl{Geochim.~Cosmochim.~Acta}}   
\def\grl{\aaref@jnl{Geophys.~Res.~Lett.}} 
\def\jcp{\aaref@jnl{J.~Chem.~Phys.}}      
\def\jgr{\aaref@jnl{J.~Geophys.~Res.}}    
\def\jqsrt{\aaref@jnl{J.~Quant.~Spec.~Radiat.~Transf.}}
\def\memsai{\aaref@jnl{Mem.~Soc.~Astron.~Italiana}}
\def\nphysa{\aaref@jnl{Nucl.~Phys.~A}}   
\def\physrep{\aaref@jnl{Phys.~Rep.}}   
\def\physscr{\aaref@jnl{Phys.~Scr}}   
\def\planss{\aaref@jnl{Planet.~Space~Sci.}}   
\def\procspie{\aaref@jnl{Proc.~SPIE}}   

\let\astap=\aap
\let\apjlett=\apjl
\let\apjsupp=\apjs
\let\applopt=\ao

\label{firstpage}
\bibliographystyle{mn2e}
\maketitle

\begin{abstract}

We aim to derive a simple analytic model to understand the essential properties of vertically settling growing dust grains in laminar protoplanetary discs. Separating the vertical dynamics from the motion  in the disc midplane, we integrate the equations of motion for both a linear and an exponential grain growth rate. Numerical integrations are performed for more complex growth models.

We find that the settling efficiency depends on the value of the dimensionless parameter $\gamma$, which characterises the relative efficiency of grain growth with respect to the gas drag. Since $\gamma$ is expected to be of the same order as the initial dust-to-gas ratio in the disc ($\simeq 10^{-2}$), grain growth enhances the energy dissipation of the dust particles and improve the settling efficiency in protoplanetary discs. This behaviour is mostly independent of the growth model considered as well as of the radial drift of the particles. 

\end{abstract}


\section{Introduction}
\label{sec:intro}

In this series of papers, we study the dynamics of growing dust grains in protoplanetary discs. In the two previous papers (Laibe et al. 2013 a,b, submitted), we have shown how grain growth interplays with the radial drift of the grains and can lead to situations where the dust particles are accreted onto the central star (the so-called radial-drift barrier) or survive in the disc. These studies assumed that the radial and the vertical motion of grains can be decoupled since they occur on very different timescales. Grains radial drift was therefore derived as if the grains motion occurred only in the disc midplane.

However, in addition to their radial evolution, grains experience a vertical motion that results from the balance between the vertical component of the central star's gravity and of the gas drag. Dust particles settle more or less efficiently to the midplane of the disc depending on their size. This motion is therefore called vertical settling. By definition, vertical settling consists of the dust motion in a laminar flow. When the disc is turbulent, the particles are stirred out of the disc midplane in a process called vertical stirring. However, turbulence is not a purely diffusive noise since turbulent fluctuations are correlated. Studying laminar flows is therefore important, as it provides the limit at infinitely large correlation times.

The vertical settling of grains with constant sizes has been studied theoretically in various papers \citep[see, e.g.,][]{Garaud2004,BF2005}. Vertical dust evolution depends essentially on the $ s/\sopt $ ratio ($\sopt$ being the optimal size of migration introduced in Paper~I) but weakly on the ratio  $\phi$ between the scale height $H$ and radius $r$, since $\phi$ is in a range of $10^{-2} -10^{-1}$ for cold protoplanetary discs. The dynamics of large grains is driven by two timescales: a typical settling time equal to the stopping time $\ts$, which characterizes the response time of a grain to the gas  drag, and a pseudo period of oscillations about the midplane which is of order $\tk$ , the Keplerian timescale. Mid-sized grains have a typical settling time of $ \tk \simeq \ts $ (on thus the settling occurs in approximately one orbit). Two time scales can also be distinguished for the vertical motion of small grains: the stopping time $\ts $ and the typical settling time $ \tk^{2} / \ts $, which increases for decreasing grain sizes. Fastest sedimentation occurs for the critical regime, i.e. for $  \tk \simeq \ts$. This regime corresponds to a typical grains size of 1 m at 1 AU in the theoretical Minimum Mass Solar Nebula model and 1 cm at 50 AU in observed Classical T-Tauri Star discs.

The vertical settling of dust is strongly affected by grain growth. In this paper, we aim to quantify the efficiency of the vertical settling of growing dust grains. Although the study is instructive and simple in comparison to the one on the radial evolution, we have not found any analytic results on the topic in the literature. We first recall the main properties of the settling of non-growing grains in Sect.~\ref{sec:settlingng}. We then generalise the harmonic oscillator approximation for non-growing grains, which we integrate analytically for both a linear and an exponential growth model, as well as numerically for other physical growth models. The results are shown to be mainly independent of the growth model considered. Moreover, the studies of both the radial and vertical motion of growing grains performed respectively in Papers~I, II  and in this paper are based on the assumptions that the motions in the disc midplane and in the vertical direction are decoupled. We test the validity of this assumption in Sect.~\ref{sec:combined} and present our conclusions in Sect. \ref{sec:conclusion}.

\section{Vertical settling of non-growing grains}
\label{sec:settlingng}

The disc is a thin, non-magnetic, non-self-graviting, inviscid perfect gas disc which is vertically isothermal. Its radial surface density and temperature are described by power-law profiles. The flow is laminar and in stationary equilibrium. Consequently, the gas velocity and density are described by well-known relations, which we presented in Paper~I. Notations are described in Appendix~\ref{App:Notations}. The equation of motion for dust grains in protoplanetary discs are given by
\begin{equation}
\left\lbrace 
\begin{array}{rcl}
\dst \frac{\mathrm{d} \tvr}{\mathrm{d} T} - \frac{\tvtheta^{2}}{R} + \frac{\tvr}{\sz}R^{-\left(p+\frac{3}{2} \right)}e^{-\frac{Z^{2}}{2R^{3-q}}} +\frac{R}{\left(R^2 + \phiz Z^{2} \right)^{3/2}} & = & 0 \\[3ex]
\dst \frac{\mathrm{d} \tvtheta}{\mathrm{d} T} + \frac{\tvtheta\tvr}{R} + \frac{\tvtheta - \sqrt{\frac{1}{R} - \etaz R^{-q} - q\left(\frac{1}{R} - \frac{1}{\sqrt{R^{2}+\phiz Z^{2}}} \right)}}{\sz}\\
\dst\times\ R^{-\left(p+\frac{3}{2} \right)}e^{-\frac{Z^{2}}{2R^{3-q}}} & = & 0 \\[2ex]
\dst \frac{\mathrm{d}^{2}Z}{\mathrm{d}T^{2}} + \frac{1}{\sz}\frac{\mathrm{d}Z}{\mathrm{d}T}R^{-\left(p+\frac{3}{2} \right)}e^{-\frac{Z^{2}}{2R^{3-q}}} + \frac{Z}{\left(R^{2} + \phiz Z^{2} \right)^{3/2}} & = & 0 .
\end{array}
\right.
\label{eq:dust3d}
\end{equation}
These equations depend on five control parameters ($\etaz$, $\sz$, $\phiz$, $p$, $q$) which are the initial dimensionless acceleration due to the pressure gradient, initial dimensionless grain size, initial disc aspect ratio and exponents of surface density and temperature profiles respectively.

\begin{figure}
\resizebox{\hsize}{!}{\includegraphics[angle=-90]{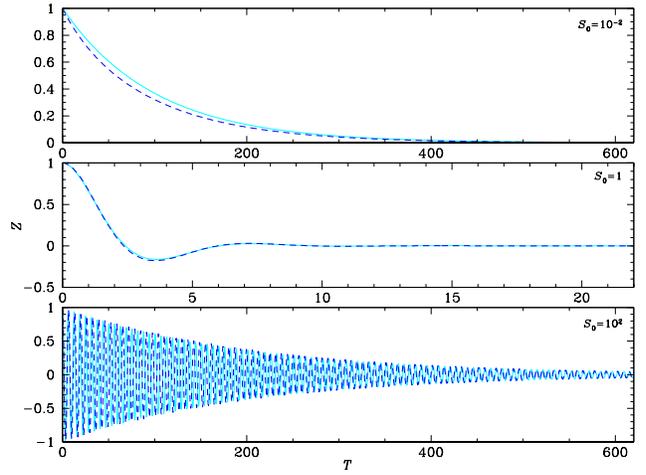}}
\caption{Vertical motion of a non-growing dust particle starting at $ Z_{0} = 1 $ with $ S_{0} = 10^{-2} $ (top), $ S_{0} = 1 $ (centre) and $ S_{0} = 10^{2} $ (bottom) obtained by numerical integration. Solid lines represent the damped harmonic oscillator motion (Eq.~\ref{eq:vertmotionadim}) and dashed lines the general case (Eq.~\ref{eq:OHgene}) for $ \phiz = 0.01$. Note the different timescale in the centre plot.} 
\label{fig:dustvert}
\end{figure}

Here we derive approximate solutions for the vertical motion of dust grains. This motion depends on the radial distance from the central star and is thus coupled to the grains radial evolution. However, its general behaviour is well-reproduced by separating both radial and vertical motions (i.e. all quantities are taken at $ r = r_{0} $, i.e. $ R = 1 $, see below for the justification).
The equation of grain dynamics on the $ \textbf{e}_{z} $ axis is given by the following Lienard equation:
\begin{equation}
\ddot{Z} + \frac{\mathrm{e}^{-\frac{Z^{2}}{2}}}{\sz} \dot{Z} + \frac{Z}{\left(1 + \phiz Z^{2} \right) ^{3/2}} = 0 .
\label{eq:OHgene}
\end{equation}
To $\mathcal{O}\left(\phiz^{2}\right)$ (for a thin disc) and $\mathcal{O}\left(Z^{2}\right)$ (for particles close to the disc midplane), or equivalently, performing a linear expansion of this equation near its fixed point $ \left(\dot{Z} = 0 , \ddot{Z} = 0 \right)  $ this becomes
\begin{equation}
\ddot{Z} + \frac{1}{\sz} \dot{Z} + Z = 0 ,
\label{eq:vertmotionadim}
\end{equation} 
which implies that near this fixed point, grain dynamics are equivalent to the damped harmonic oscillator. Fig.~\ref{fig:dustvert} compares the vertical motion of the dust particles  given the harmonic oscillator approximation (Eq.~\ref{eq:vertmotionadim}) and the general case (Eq.~\ref{eq:OHgene}) and show that the harmonic oscillator approximation is justified. Moreover, given the sizes $\sz$ of the dust particles, the three characteristic regimes of the damped harmonic oscillator can then be distinguished:

\begin{enumerate}

\item $ \sz > 1/2 $: underdamped oscillator

\begin{equation}
\left\lbrace
\begin{array}{rcl}
Z(T) & = & e^{-\frac{1}{2 \sz} T} \left[ A \cos(\lambda T) + B \sin(\lambda T) \right] \\[2ex]
\lambda & = & \dst\sqrt{1 - \left( \frac{1}{2 \sz}\right) ^{2}} \\[2ex]
A & = & Z(0) \\[2ex]
B & = & \dst\frac{1}{\lambda} \left(\frac{1}{2 \sz} Z \left( 0 \right)  + \dot{Z} \left( 0 \right) \right) . 
\end{array}
\right.
\label{eq:underdamped}
\end{equation}

\item $ \sz = 1/2 $: critical regime

\begin{equation}
\left\lbrace
\begin{array}{rcl}
Z \left( T\right) & = & \left( A + B T\right) e^{- T} \\[2ex]
A & = & Z\left( 0 \right) \\[2ex]
B & = & Z\left( 0 \right) + \dot{Z} \left( 0 \right)  . 
\end{array}
\right.
\label{eq:criticaldamped}
\end{equation}

\item $ \sz < 1/2 $: overdamped oscillator

\begin{equation}
\left\lbrace
\begin{array}{rcl}
Z \left( T\right) & = &  A e^{\lambda_{+} T} + B e^{\lambda_{-} T}\\[2ex]
\lambda_{\pm} & = & - \dst\frac{1}{2 \sz} \pm \sqrt{\left( \frac{1}{2 \sz}\right) ^{2} - 1  } \\[2ex]
A & = & Z\dst\left( 0 \right) + \frac{\lambda_{+} Z\left( 0 \right) - \dot{Z} \left( 0 \right) }{ \lambda_{-} - \lambda_{+} } \\[2ex]
B & = & - \dst\frac{\lambda_{+} Z\left( 0 \right) - \dot{Z} \left( 0 \right) }{ \lambda_{-} - \lambda_{+} }  . 
\end{array}
\right.
\label{eq:overdamped}
\end{equation}
\end{enumerate}

The vertical settling of a grain occurs on a dimensionless timescale that is shorter than the migration timescale. From Eq.~\ref{eq:underdamped} (resp. Eq.~\ref{eq:overdamped}), for small (resp. large) grains, the dimensionless settling time is $1/S$ (resp. $S$). The fastest sedimentation occurs for the critical regime, i.e. for $S = 1/2$ (Eq.~\ref{eq:criticaldamped}). A good approximation for the typical settling time $T_{\rm set}$ is therefore
\begin{equation}
T_{\rm set} \simeq \frac{1 + S^{2}}{S} .
\label{eq:tset}
\end{equation}
The typical migration time is obtained from Eq.~16 of \citet{Laibe2012} estimating $\dst R \left/ \frac{\mathrm{d}R}{\mathrm{d}T} \right.$ at $R=1$, providing
\begin{equation}
T_{\rm mig} \simeq \frac{1 + S^{2}}{\etaz S} ,
\label{eq:tset}
\end{equation}
$T_{\rm set} / T_{\rm mig}  = \mathcal{O}(\etaz) \simeq 10^{-2}$, settling is almost a hundred times faster than migration and both the vertical and the radial motion are decoupled \citep{Garaud2004}. 
\section{Vertical settling of growing grains}
\label{sec:settling}

\subsection{Growth models}
\label{sec:growth}

In the case of growing grains, the ratio $\frac{\ts}{t_{\mathrm{k}}}$ of the drag ad the Keplerian timescales evolves, changing the vertical evolution of the particles. Following the same format as in papers~I and II and the notations in Appendix~\ref{App:Notations}, we have%
\begin{equation}
\St (T) =  \frac{\ts}{t_{\mathrm{k}}} = 
\frac{s}{\frac{\rhog \cs}{\rhod \Omega_{\mathrm{k}}}} = \frac{s}{s_{\mathrm{opt}}} =
S(T) R^{p}e^{\frac{Z^{2}}{2R^{3-q}}},
\label{eq:defsz}
\end{equation}
where $ \Omega_{\mathrm{k}}$ is the Keplerian frequency and $\cs$, $\rhog$, $\rhod$ the gas sound speed, the gas density and the dust density respectively. As with the settling of non-growing grains, we assume first that the vertical motion occurs much faster than the radial motion, implying that $R=1$ during the grain's settling and secondly, we assume that the oscillations are sufficiently close to the disc midplane ($Z \ll 1$) during the evolution. For non growing grains, we have seen that these assumptions have a negligible impact on the vertical evolution. Thus, Eq,~\ref{eq:defsz} reduces to 
\begin{equation}
\St (T)  = S (T) .
\label{eq:defsz_simp}
\end{equation}
Substituting $\sz$ by $S$ in Eq.~\ref{eq:vertmotionadim} provides the differential equation which governs the vertical motion of the grains :
\begin{equation}
\ddot{Z} + \frac{1}{S\left( T\right) } \dot{Z} + Z = 0 .
\label{eq:genericvertgr}
\end{equation}
The evolution of $S(T)$ is governed by the growth rate of the particles. Several models of grain growth have been introduced and studied in Paper~II. Importantly, it is explained that for a cold disc at $Z \ll 1$ the growth rate of the particles is of the form:
\begin{equation}
\frac{\mathrm{d}S}{\mathrm{d}T} = \gamma f(S) ,
\label{eq:grpart}
\end{equation}
where $f$ is a function of the grain size which depends on the models for the relative turbulent velocities between the particles and the scale height of the dust layer considered. As discussed in Paper~II, $\gamma$ is of order $\epsilon_{0}$, the initial dust-to-gas ratio of the disc, which is $ \simeq 10^{-2}$ in protoplanetary discs. With the most recent models of dust and gas turbulence modelling (see Paper~II for a discussion), $f$ is of the form
\begin{equation}
f(S) = \frac{S}{1+S} \simeq S^{y_{\rm g}} ,
\label{eq:f_form}
\end{equation}
with $y_{\rm g}  = 1$ for $S \ll 1 $ and $y_{\rm g} = 0$ for $S \gg 1$ . $f$ often reduces to a simple power law of exponent $y_{\rm g}$ when treating the small and the large grains separately. In this case, as discussed in Paper~II, $y_{\rm g}$ can take values of order unity in the case of realistic growth rates and differs from the case $S \ll 1 $to the case $S \gg 1 $. The size evolution is thus given by
\begin{equation}
S(T) = \left( \left( -y_{\rm g} + 1 \right) \gamma T + \sz^{-y_{\rm g}+1}\right)^{\frac{1}{-y_{\rm g} + 1}} ,
\label{eq:size_gene}
\end{equation}
if $y_{\rm g} \ne 1$ and
\begin{equation}
S(T) = \sz \rm{e}^{\gamma T} ,
\label{eq:size_one}
\end{equation}
if $y_{\rm g} = 1$. The case $y_{\rm g} = 0$ (linear growth, Eq.~19 of Paper~I) corresponds to the limit of the large grains in Eq.~\ref{eq:f_form} and the case $y_{\rm g} = 1$ to the limit of the small grains. It is is also straightforward to derive the general expression of the size evolution to the power-law toy model (Eq.~23) used in Paper~I.

\subsection{Linear growth model}
\label{sec:analyticvert}

\begin{figure}
\resizebox{\hsize}{!}{\includegraphics[angle=-90]{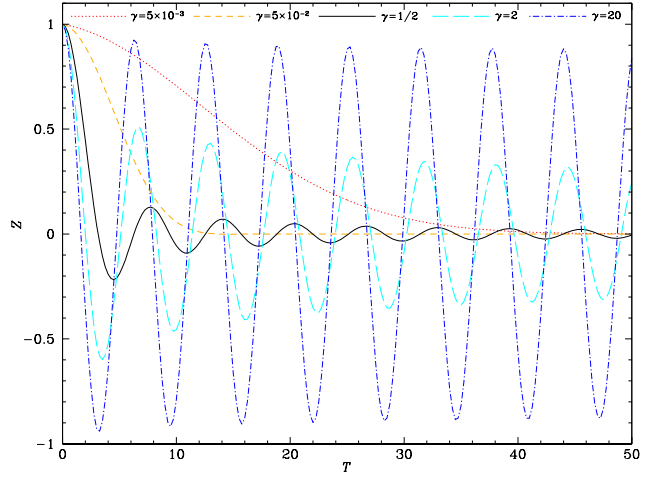}}
\caption{Vertical motion of a growing dust particle with time T starting at $ Z_0 = 1 $ with $ \sz = 10^{-2} $ and $ \gamma = 5\times10^{-3}$, $5\times10^{-2}$, $0.5$, 2 and 20. The most efficient settling is obtained for $\gamma = 1/2$.}
\label{fig:variousdelta}
\end{figure}

We investigate the coupling between the growth and the settling using the simplest linear growth model from Eq.~\ref{eq:size_gene} with $y_{\rm g} = 0$ giving
\begin{equation}
\ddot{Z} + \frac{1}{ \sz + \gamma T } \dot{Z} + Z = 0 .
\label{eq:growthlinoscill}
\end{equation}
To solve this differential equation, we introduce the auxiliary function $ \varsigma\left( T\right) $ such that
\begin{equation}
Z\left( T\right) = \varsigma \left( T\right) \times e^{\dst-\frac{1}{2} \int_{0}^{T} \frac{\ddiff{\tau}}{\sz + \gamma \tau} } = \varsigma \left( T\right) \times \left( 1 + \frac{\gamma T}{\sz}\right) ^{-\frac{1}{2 \gamma}} .
\label{eq:solveaux}
\end{equation}
 Hence, $ Z\left( T \right)  $ is the product of two functions: $ \left( 1 + \frac{\gamma T}{\sz}\right) ^{-\frac{1}{2 \gamma}} $ and a function $\varsigma$ which satisfies
\begin{equation}
\ddot{\varsigma} + I\left( T\right) \varsigma  = 0 ,
\label{eq:equaux}
\end{equation}
with
\begin{equation}
I\left( T\right) = 1 - \frac{1}{4} \frac{1 - 2 \gamma}{\left( \sz + \gamma T \right) ^{2}} .
\label{eq:Itau}
\end{equation}
The general solution of Eq. (\ref{eq:equaux}) with (\ref{eq:Itau}) is
\begin{equation}
\begin{array}{rcl}
\varsigma \left( T\right) & = & \dst\ind{C}{1} \sqrt{\sz + \gamma T } \,\mathcal{J}_{\frac{1}{2} \frac{\gamma - 1}{\gamma}} \left( \frac{\sz + \gamma T}{\gamma}\right)  \\[2ex]
& & +\ \dst\ind{C}{2} \sqrt{\sz + \gamma T } \,\mathcal{Y}_{\frac{1}{2} \frac{\gamma - 1}{\gamma}} \left( \frac{\sz + \gamma T}{\gamma}\right) ,
\end{array}
\label{eq:solveu}
\end{equation}
where $ \mathcal{J}_{\nu} $ and $ \mathcal{Y}_{\nu} $ are the Bessel functions of first and second kind of order $ \nu $, and $ \ind{C}{1}  $, $ \ind{C}{2}  $ are constants determined by the initial conditions. Therefore, the solution of Eq. (\ref{eq:growthlinoscill}) is
\begin{equation}
\begin{array}{rcl}
Z\left( T\right) & = & \dst\left( 1 + \frac{\gamma T}{\sz}\right) ^{-\frac{1}{2 \gamma}} \left[ \ind{C}{1} \sqrt{\sz + \gamma T } \,\mathcal{J}_{\frac{1}{2} \frac{\gamma - 1}{\gamma}} \left( \frac{\sz + \gamma T}{\gamma}\right) \right. \\[2ex]
 &&+\dst\left. \ind{C}{2} \sqrt{\sz + \gamma T } \,\mathcal{Y}_{\frac{1}{2} \frac{\gamma - 1}{\gamma}} \left( \frac{\sz + \gamma T}{\gamma}\right)  \right]  .
\end{array}
\label{eq:solveZaux}
\end{equation}
If $ Z\left( T = 0\right) = \ind{Z}{0}  $ and $ \dot{Z} \left( T = 0\right) = 0 $, the constants $ \ind{C}{1}  $ and $ \ind{C}{2}  $ are given by
\begin{eqnarray}
\ind{C}{1} &=& -\frac{\ind{Z}{0}}{\sz^{3/2} } \frac{\left( \gamma - 1\right) \mathcal{Y}_{\frac{\gamma - 1}{2 \gamma}} \left( \frac{\sz}{\gamma}\right) - \sz \mathcal{Y}_{\frac{3 \gamma - 1}{2 \gamma}} \left( \frac{\sz}{\gamma}\right)     }{\mathcal{Y}_{\frac{3 \gamma - 1}{2 \gamma}} \left( \frac{\sz}{\gamma}\right) \mathcal{J}_{\frac{\gamma - 1}{2 \gamma}} \left( \frac{\sz}{\gamma}\right)  - \mathcal{Y}_{\frac{\gamma - 1}{2 \gamma}} \left( \frac{\sz}{\gamma}\right) \mathcal{J}_{\frac{3 \gamma - 1}{2 \gamma}} \left( \frac{\sz}{\gamma}\right)  } ,\\
\label{constc1}
\ind{C}{2} &=& \frac{\ind{Z}{0}}{\sz^{3/2} } \frac{\left( \gamma - 1\right) \mathcal{J}_{\frac{\gamma - 1}{2 \gamma}} \left( \frac{\sz}{\gamma}\right) - \sz \mathcal{J}_{\frac{3 \gamma - 1}{2 \gamma}} \left( \frac{\sz}{\gamma}\right)     }{\mathcal{Y}_{\frac{3 \gamma - 1}{2 \gamma}} \left( \frac{\sz}{\gamma}\right) \mathcal{J}_{\frac{\gamma - 1}{2 \gamma}} \left( \frac{\sz}{\gamma}\right)  - \mathcal{Y}_{\frac{\gamma - 1}{2 \gamma}} \left( \frac{\sz}{\gamma}\right) \mathcal{J}_{\frac{3 \gamma - 1}{2 \gamma}} \left( \frac{\sz}{2 \gamma}\right)  } .
\label{eq:constc2}
\end{eqnarray}
The sign of the function $ I\left( T\right) $, for which we have $\mathrm{d}I/\mathrm{d}T = \frac{1}{2} \gamma \left( 1 - 2 \gamma \right) / \left( \sz + \gamma T\right) ^{3}$, provides information on the oscillating behaviour of the solution. Thus three cases can be distinguished, separated by the critical value for settling $\gammacs=\frac{1}{2}$:

\begin{enumerate}

\item{$0<\gamma<\gammacs=\frac{1}{2}$ and $\mathrm{d}I/\mathrm{d}T>0$:}

$I$ increases from its initial value $I(T=0)<1$ and \mbox{$\lim\limits_{\substack{T \to +\infty}} I = 1$}. If $\sz > \frac{1}{2}\sqrt{1-2\gamma}$, then $ I\left( T = 0\right) > 0 $ and $I$ is positive at all times: the solution is always pseudo-oscillating. The dust particle is always decoupled from the gas as the size can only increase and the dust evolution follows the large grain regime.  This is of minor interest in the context of growing grains. We therefore consider the interesting case $\sz < \frac{1}{2}\sqrt{1-2\gamma}$, for which $ I\left( T = 0\right) < 0 $ and $ I $ becomes positive for $ T > \frac{\frac{1}{2}\sqrt{1-2\gamma} - \sz }{\gamma}$: the solution transitions from a monotonic decay to a pseudo-oscillating regime, indicating that particles decouple from the gas.

\item{$\gamma=\gammacs=\frac{1}{2}$ and $\mathrm{d}I/\mathrm{d}T=0$:}

In this limiting case, $ I\left( T\right) =1 $ for all time. Phase lag due to damped oscillations is exactly counterbalanced by the decrease of the drag caused by grain growth.

\item{$\gamma>\gammacs=\frac{1}{2}$ and $\mathrm{d}I/\mathrm{d}T<0$:}

$I$ decreases from its initial value $I(T=0)>1$ and $\lim\limits_{\substack{T \to +\infty}} I = 1$. Since $I$ is always greater than 1, the solution is pseudo-oscillating at a frequency larger than the Keplerian frequency.

\end{enumerate}

\begin{figure}
\resizebox{\hsize}{!}{\includegraphics{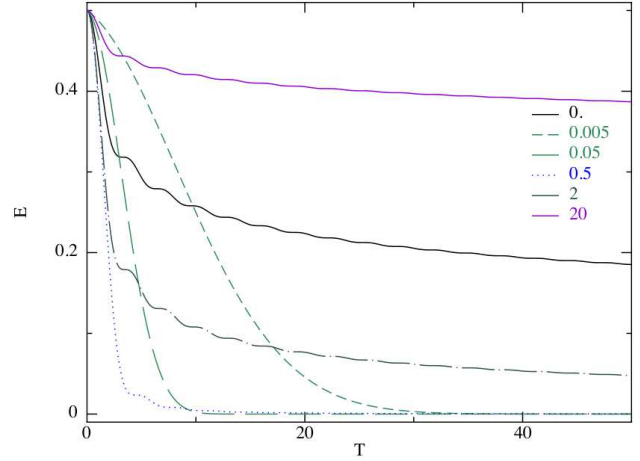}}
\caption{Evolution of the total energy for dust particles starting at $ Z_0 = 1 $ with $ \sz = 10^{-2} $ and $ \gamma = 0$ (no growth), $5\times10^{-3}$, $5\times10^{-2}$, $1/2$, 2 and 20. The most efficient dissipation corresponds to the most efficient settling obtained for $\gamma = 1/2$.}
\label{fig:enlin}
\end{figure}

The envelope of the solution, which determines the damping of the dust's vertical motion, is given by the product of $ \left( 1 + \frac{\gamma T}{\sz}\right) ^{-\frac{1}{2 \gamma}} $, $\sqrt{\sz + \gamma T }$ and the envelope of the Bessel functions. While not  transparent, it is qualitatively interpretable. The dust behaviour for different values of $\gamma$ are shown in Fig.~\ref{fig:variousdelta} for $\sz=10^{-2}$. We focus on initially small grains ($\sz \ll 1$) because they correspond to the grains which originate in the interstellar medium and are involved in planet formation. In the case of the slow growth regime ($ \gamma < 1/2 $), the vertical dust motion is damped efficiently: particles settle to the midplane of the disc before they have time to grow and decouple from the gas. However, as grains decouple slowly from the gas as they settle, drag becomes weaker. Thus dust settling occurs faster than for non-growing grains and the settling rate increases for increasing values of $\gamma$. On the contrary, in the fast growth regime ($ \gamma >1/2 $), dust particles grow fast enough to decouple from the gas before they feel the gas drag and their settling timescale becomes much longer. In this case, the settling time of the grain increases dramatically with $\gamma$ since an asymptotic expansion of Eq.~\ref{eq:solveZaux} for $\gamma \gg 1$ provides $T_{\rm sett} = \mathcal{O}(\mathrm{e}^{\gamma}/\gamma)$. In the intermediate case ($ \gamma = 1/2 $), dust particles grow in the same timescale as they settle to the midplane where they decouple from the gas. This corresponds to the most efficient regime of settling.

Additionally, Fig.~\ref{fig:enlin} shows the evolution of  the total energy $E$ given by
\begin{equation}
E = \frac{1}{2} \left(\dot{Z}^{2} + Z^{2} \right) = E(T=0) - \int_{0}^{T} \frac{\dot{Z}^{2}}{S(T')} \mathrm{d}T' ,
\label{eq:toten}
\end{equation}
for different values of the growth parameter $\gamma$. Even small values of $\gamma$ (e.g. $ 5\times10^{-3}$ or $ 5\times10^{-2}$ which correspond to real protoplanetary discs) provide a dissipation which is more efficient than for the case without any growth.  The most efficient dissipation corresponds to the most efficient settling obtained for $\gamma = 1/2$. Increasing again the value of $\gamma$ leads to a less efficient dissipation process. In the limit of large values of $\gamma$, the particles decouple so quickly from the gas that the dissipation is even less efficient than for the case without any growth.

\subsection{Other growth models}
\label{app:appendixexp}

\begin{figure}
\resizebox{\hsize}{!}{\includegraphics{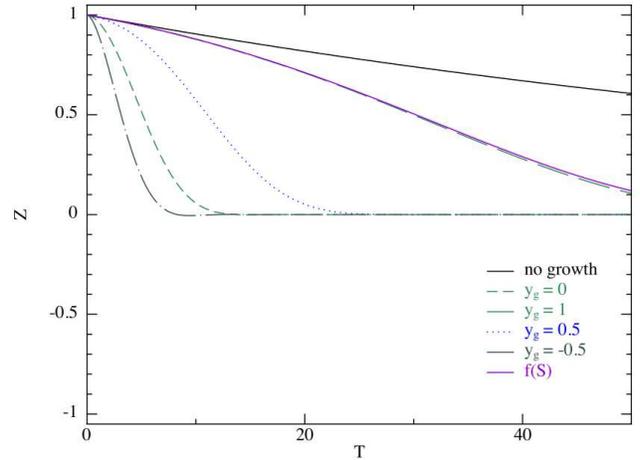}}
\caption{Vertical motion of a growing dust particle with starting at $ Z_0 = 1 $ with $ \sz = 10^{-2} $ and $ \gamma = 5\times10^{-2}$ for different growth models.}
\label{fig:eps5dm2}
\end{figure}

We can also integrate the vertical motion of the grains for several growth models: specifically power laws with $y_{\rm g} = 1, -0.5, 0.5$ and growth rate given by the function $f$ of Eq.~\ref{eq:f_form}. For the exponential growth rate model ($y_{\rm g} = 1$), we derive analytically the evolution of the dimensionless vertical coordinate, which is given by

\begin{equation}
\begin{array}{r}
Z\left( T \right)  = \left[ B_{1} e^{\frac{\pi}{\gamma}} \left( \gamma \sz e^{\gamma t}\right)^{\frac{i}{\gamma}} M\left( -\frac{i}{\gamma}, \frac{-2i+\gamma}{\gamma} , \frac{e^{-\gamma t}}{\gamma \sz} \right) \right. \\[2ex]
+ \left. B_{2} e^{\frac{-\pi}{\gamma}}  \left( \gamma \sz e^{\gamma t}\right)^{-\frac{i}{\gamma}}  M\left( \frac{i}{\gamma}, \frac{2i+\gamma}{\gamma} , \frac{e^{-\gamma t}}{\gamma \sz} \right) \right) , 
\end{array}
\label{eq:exprZexpo}
\end{equation}
where $i^{2} = -1$, $B_{1,2}$ are constants which are determined by the initial conditions, and $M(a,b,z)$ is the M-Kummer confluent hypergeometic function of indices $a$ and $b$ with respect to $z$. For the other growth models, we did not manage to derive the evolution analytically and therefore we must integrate the equations numerically.


Fig.~\ref{fig:eps5dm2} shows the vertical behaviour of the particle with $\gamma = 5\times 10^{-2}$ (similar plots with $\gamma = 0.5$ and $\gamma = 5$ are shown in Appendix~\ref{app:other}).

Smaller values of $y_{\rm g}$ provide a more efficient settling of the particles because small grains grow faster. Moreover, the exponential growth provides a good approximation of the function $f$ in the case of small values of $\gamma$ (particles settle having mainly $S < 1$), whereas the linear growth provides a good approximation of the function $f$ in the case of large values of $\gamma$ (particles settle having mainly $S > 1$). However, overall, the nature of settling for the different growth models are very similar. In particular, for $\gamma = 5\times 10^{-2}$, the grains settle much faster than in the case without any growth. Therefore, the conclusion that grain growth enhances vertical settling's efficiency in protoplanetary discs hold whatever the growth model considered.

\section{Combining the radial and the vertical motion}
\label{sec:combined}

\begin{figure}
\resizebox{\hsize}{!}{\includegraphics{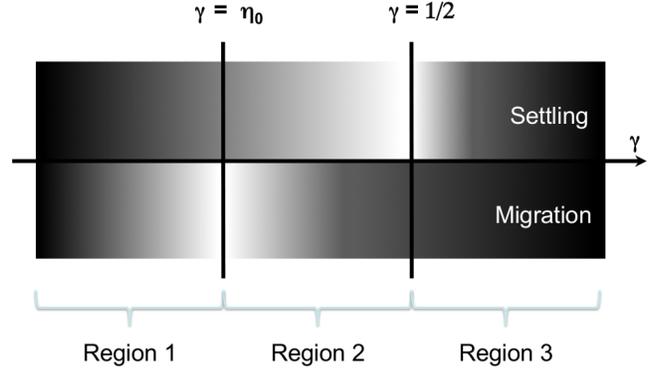}}
\caption{Illustration of the relative efficiency of the vertical and radial behaviour of growing dust grains for different values of the growth rate $\gamma$, all the other parameters being fixed. The parameter space is divided in three regions delimited by the white zones indicating an optimal dynamics for the vertical and the radial motion.} 
\label{fig:comportdust}
\end{figure}

In the studies performed in Papers~I, II and in this Paper, we found two interesting values of the growth rate $\gamma$: $ \gamma = \etaz$ (giving $\Lambda =1$) and $\gamma = 1/2$, corresponding respectively to the optimal values of $\gamma$ for the migration and the vertical settling process.
The parameter space can therefore be divided in three regions as shown in Fig.~\ref{fig:comportdust}: $\gamma < \etaz$ (region 1), $\etaz < \gamma < 1/2$ (region 2) and $\gamma > 1/2$ (region 3), noting that $ \mathcal{O}\left(\etaz \right)<1/2$ for real discs. In each of these regions, the efficiency of the vertical (resp. radial) motion is represented by the brightness of the top (resp. bottom) colour bar. Importantly, this plot provides an indication of the relative efficiencies of the migration and settling processes, all the other parameters being fixed. It does not, however, predict the grains final state (decoupling at a finite radius, pile-up or accretion onto the star) but support the hypothesis of the decoupling between the radial and the vertical motions. 

\begin{figure}
\resizebox{\hsize}{!}{\includegraphics{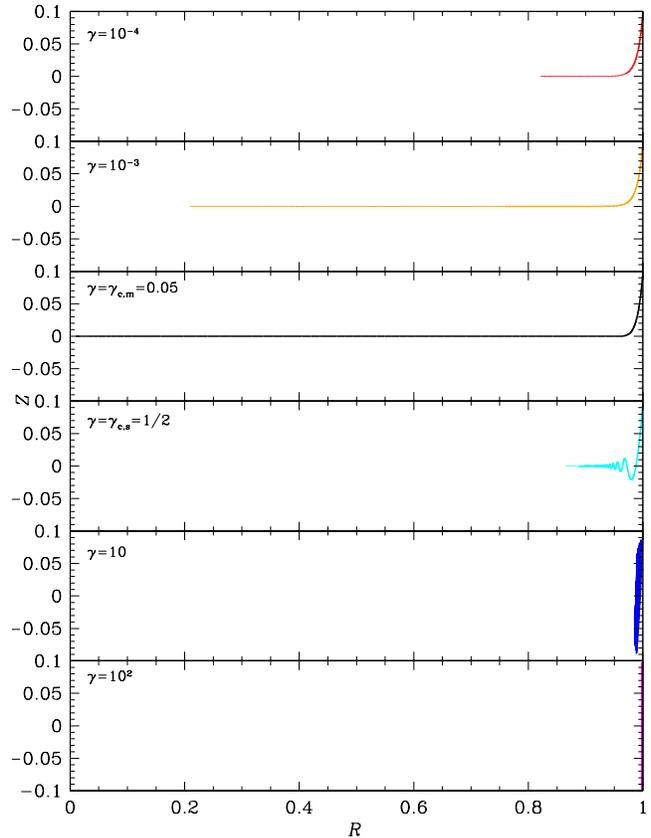}}
\caption{Trajectories in the $(R,Z)$ plane of dust grains starting at ($R=1, Z=0.1$) after a time $T=240$ (thick lines) and $T=628$ (thin lines) with $ \sz = 10^{-3} $, $ \etaz = 10^{-2} $, $p=3/2$, $q=3/4$, $n=1$  and $ \gamma = 10^{-4},10^{-3}, 5\times10^{-2}, 1/2, 10,10^{2} $  (from top to bottom). }
\label{dustgrowthradvert}
\end{figure}

Until now we have assumed that the grains radial and vertical motion are decoupled. We test this hypothesis by numerically integrating the combined vertical and radial equations of motion (Eq.~\ref{eq:dust3d}) in three dimensions, substituting $\sz$ by $S(T)$. The resulting trajectories in the $RZ$-plane for $\gamma = 10^{-4},10^{-3}, 5\times10^{-2}$, 1/2, 10, and 100 are shown in Fig.~\ref{dustgrowthradvert}. We studied the case of $\etaz = 10^{-2}$ with $p=3/2$, $q=3/4$, $n=1$ and $\sz = 10^{-3}$. We find that the most efficient value of $\gamma$ to reach $R_\mathrm{f} = 0.01$ is  $\gammacm = 0.050$, which corresponds to a time $T_\mathrm{f} = 240$.

\begin{figure}
\resizebox{\hsize}{!}{\includegraphics{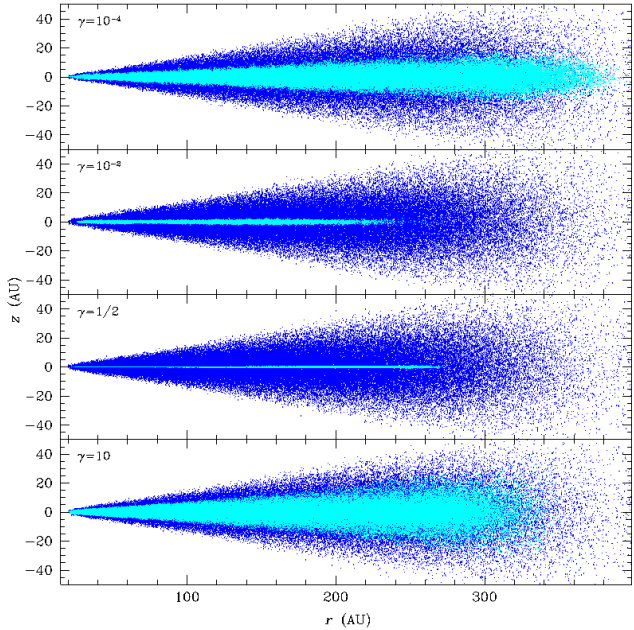}}
\caption{Radial grain size distribution obtained with our SPH code after $10^{5}$ years for $\gamma=10^{-4}$, $10^{-2}$, $\frac{1}{2}$ and 10. Dark blue: gas. Light blue: dust. Thinner disc distributions are obtained when the ratio between the growth time scale is the same as the optimal settling timescale ($\gamma = 1/2$).} 
\label{fig:Elisdistrib}
\end{figure}

\begin{figure}
\resizebox{\hsize}{!}{\includegraphics[angle=-90]{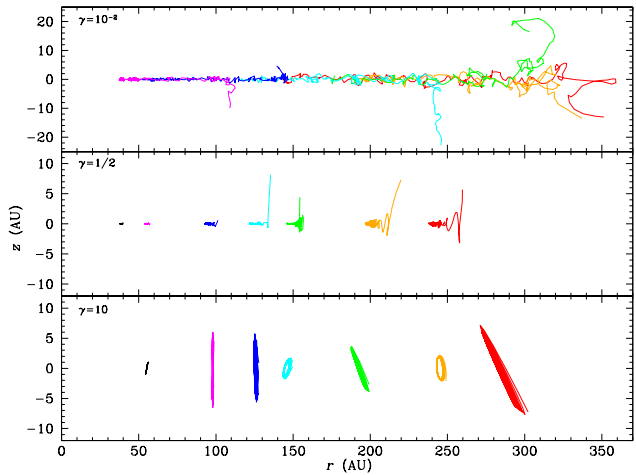}}
\caption{Trajectories in the $(r,z)$ plane of individual SPH dust particles for $\gamma=10^{-2}$, $\frac{1}{2}$ and 10. Top: the migration is highly efficient (the growth timescale is of order the migration timescale), contrary to the settling. Centre: the settling efficiency is optimal (the growth timescale of order the optimal settling timescale), but the migration is no longer efficient. Bottom: Particles are almost instantaneously decoupled from the gas and are Keplerian orbits.}
\label{fig:Elistraj}
\end{figure}

In region 1of Fig~\ref{fig:comportdust}, particles with small growth rates ($\gamma<\gammacm$) settle to the midplane and then migrate towards the central star. Growth is not efficient enough to make the particles decouple from the gas before they reach the inner region of the disc. The migration efficiency is optimal for $\gamma=\gammacm$. In region 2 of Fig~\ref{fig:comportdust}, grains with intermediate growth rates ($\gammacm<\gamma<\gammacs=1/2$) grow as they settle to the midplane, radially migrate, but they decouple from the gas before reaching the central star and therefore experience an extremely slow migration motion. When $\gamma=\gammacs=1/2$ (the border of regions 2 and 3 of Fig~\ref{fig:comportdust}), the growth is optimal and particles decouple from the gas just as they reach the midplane. Thus particles migrate slightly while efficiently settling to the midplane (the envelope of vertical oscillations decreases very quickly). In region 3 of Fig~\ref{fig:comportdust}, the larger growth rates ($\gamma>\gammacs=1/2$) ensure that particles grow very efficiently and rapidly decouple from the gas. They do not settle to the midplane (as there is no gas damping once they decouple from the gas phase) and they experience a very small migration motion. In all cases, the vertical motion occurs much faster than the radial motion and the predictions done assuming that both motions are decoupled hold.

\begin{figure}
\resizebox{\hsize}{!}{\includegraphics{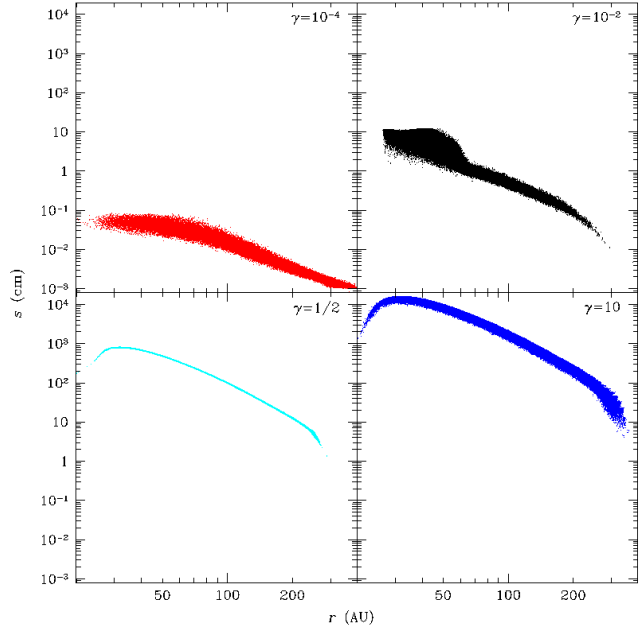}}
\caption{Radial grain size distribution obtained with our SPH code after $10^{5}$ years for $\gamma=10^{-4}$, $10^{-2}$, $\frac{1}{2}$ and 10. Distinct shapes are seen in the grains size distributions, with the thickness of the profile related to the vertical spreading of the grains.  For $\gamma = 10^{-4}$, particles near the inner edge are accreted. For $\gamma = 10^{-2}$, the migration is more efficient and particles that were initially far from the inner edge are piling up. For $\gamma = 1/2$, the settling is optimal (hence the thin profile) but migration is not efficient. For $\gamma = 10$, particles decouple almost instantaneously and follow the initial gas distribution. } 
\label{fig:Elissize}
\end{figure}

To pedagogically illustrate the effect of the linear growth rate discussed in Sect.~\ref{sec:analyticvert} on the resulting dust distribution in protoplanetary discs, we also run simple simulations with the 3D two-phase (gas and dust) Smoothed Particles Hydrodynamics (SPH) code described in \citet{BF2005} with an initial setup similar to the one described in \citet{Laibe2008}. We start with a uniform grain size $s_{0} = 10$~$\mu$m (which corresponds to $ S_{0} = 10^{-3} $ at 50~AU) in a disc where $\etaz = 10^{-2}$ at 5 AU.  The different evolutionary regimes predicted by the constant growth rate model and summarised in Fig.~\ref{fig:comportdust} are seen in the radial dust distribution in edge-on views of the disc shown in Fig.~\ref{fig:Elisdistrib}. For $ \gamma = 10^{-4}$, particles experience only weak vertical settling which is characteristic of small grains. Their migrations rate is also slow and thus they remain radially extended throughout the disk. For $ \gamma = 10^{-2} $,  both the vertical settling and the radial migration are very efficient. The particles close to the inner disc edge are rapidly accreted. However, for particles  far  from the inner edge, the pile-up is efficient enough to strongly retard the inward motion. For $\gamma = 1/2 $, particles settle very efficiently to the disc midplane (minimum disc thickness). The less efficient migration also provides a larger disc. Grains which are not near the disc inner edge  experience a slow migration regime and are not  depleted. For $ \gamma > 1/2 $, growth occurs very rapidly: grains decouple from the gas in a fraction of an orbit and effectively remain on fixed Keplerian orbits and  hence are distributed over the entire disc.

These different dust behaviours can also be illustrated by using the Lagrangian property of the SPH formalism and plotting trajectories of SPH dust particles for different values of $ \gamma $ in the $RZ$-plane (see Fig.~\ref{fig:Elistraj}). For $\gamma=10^{-2}$, we see the particles fall to the midplane and migrate radially.  When $\gamma=\frac{1}{2}$, the vertical settling is very efficient, but the migration is not and the values of $p$ and $q$ cause the dust to pile up radially.  When $ \gamma > \frac{1}{2}$, the migration is completely inefficient for the grains to reach the inner discs regions.  When $ \gamma = 10 $, particles rapidly decouple from the gas and remain on inclined elliptical Keplarian orbits and can be seen oscillating about the disc midplane.

We also plot the dust size distribution obtained after approximately $10^{5}$ years in Fig.~\ref{fig:Elissize}. With our initial choice of $(p,q) = \left( \frac{3}{2},\frac{3}{4}\right)$ this ensures that $-p+q+\frac{1}{2}<0$ and thus grains remain in the disc. Note however that the SPH disc is truncated at $r_{\rm in}=20$~AU for reasons of computational efficiency. Thus grains can be lost from the simulation if the time taken to reach $r_{\rm in}$ is smaller than the simulation duration. For $ \gamma = 10^{-4} $, particles close to the inner edge are lost, whereas for $ \gamma = 10^{-2} $ the grains pile-up in the inner disc regions, staying beyond $\sim 25$~AU. For the largest values of $ \gamma $, particles decouple very quickly from the gas and remain distributed throughout the disc. The thickness of the size distribution corresponds to the dependency of $ \sopt $ with respect to $ z $ due to the vertical density profile. This is smallest for $\gamma=\gammacs=1/2$, for which vertical settling is most efficient and the dust disc is thinnest. As a conclusion, we find that the assumption of decoupling the radial and the vertical motion of the grain is verified both in our direct integration of the equations of motion (see Fig.~\ref{dustgrowthradvert}) and in our SPH simulations (see Fig.~\ref{fig:Elistraj}).

\section{Conclusions and perspectives}
\label{sec:conclusion}

In this paper, we have studied the vertical settling of growing dust grains in protoplanetary discs, using different rates for the grain growth and integrating the equations of evolution both analytically and numerically. The main results of the study are:

\begin{enumerate}

\item The vertical motion of growing dust grains is governed by the value of the dimensionless parameter which represents the relative efficiency between the growth and the drag, which we denote $\gamma$. In protoplanetary discs, $\gamma$ is of order $\epsilon_{0}$, the initial dust-to-gas ratio of the disc, which is of order $10^{-2}$. This implies that the the growth is not too efficient  (the effective $\gamma$ being much smaller than the critical value of $1/2$) enabling particles to settle toward the disc midplane where they concentrate.

\item All the growth models we have tested give essentially the same behaviour as the linear growth model. We therefore suggest that the results of this study are generalizable and that the solution we derive provides a good analytic prescription for the vertical evolution of the particles.  

\item  Simultaneously integrating both the radial and the vertical motion of the particles shows that the vertical settling of the particles occurs much faster than the radial drift of the particles, justifying the assumption of separating the radial drift and the vertical settling. This is a standard and well-know result for non-growing grains (see e.g. \citealt{Garaud2004}), we have shown it also holds for growing grains.

\item Combining the results for the radial drift with the study of the vertical settling of dust grains, we distinguish three major regimes for growing grains: $\gamma < \eta_{0}$, $\eta_{0}< \gamma < 1/2$, $1/2 < \gamma$, the first two being the most relevant for the context of planet formation. Initially small grains grow as they settle to the midplane, the settling motion being faster than for non-growing grains. Varying $\gamma$ results in distinct profiles for the grain size distribution as well as their spatial distributions.

\end{enumerate}

Importantly, dust concentration of growing grains in the disc mid-plane has been proven to occur when the disc is laminar. If not, turbulent fluctuations from the gas may spread the dust particles out of the disc mid-plane. This vertical stirring is widely supposed to prevent grains to concentrate in the disc mid-plane and form planet by gravitational instability for non growing grains. This issue will be addressed in the case of growing dust grains in a forthcoming paper.

\section*{acknowledgements}
This research was partially supported by the Programme National de Physique
Stellaire and the Programme National de Plan\'etologie of CNRS/INSU, France,
and the Agence Nationale de la Recherche (ANR) of France through contract
ANR-07-BLAN-0221. STM acknowledges the support of a Swinburne Special Studies Program. GL is grateful to the Australian Research Council for funding via Discovery project grant DP1094585 and acknowledges funding from the European Research Council for the FP7 ERC advanced grant project ECOGAL. JFG's research was conducted within the Lyon Institute of Origins under grant ANR-10-LABX-66

\bibliography{bibliodelta}

\begin{appendix}
\section{Notations}
\label{App:Notations}

The notations and conventions used throughout this paper are summarized in Table~\ref{tabnote}.
\begin{table}
\begin{center}
\begin{tabular}{ll}
\hline Symbol & Meaning \\ \hline
$M$ & Mass of the central star \\
$\textbf{g}$ & Gravity field of the central star \\ 
$\Rz$ & Initial distance to the central star \\
$\rhog$ & Gas density \\
$\brhog\left(r\right)$ & $\rhog\left(r,z=0\right)$ \\
$\cs$ & Gas sound speed \\
$\bcs\left(r\right)$ & $\cs\left(r,z=0\right)$ \\
$\csz$ & Gas sound speed at $\Rz$ \\
$T$ & Dimensionless time\\
$\mathcal{T}$ & Gas temperature ($\mathcal{T}_{0}$: value at $\Rz$)\\ 
$\Sigmaz$ & Gas surface density at $\Rz$ \\
$p$ & Radial surface density exponent \\
$q$ & Radial temperature exponent \\
$P$ & Gas pressure \\
$v_{\mathrm{k}}$ & Keplerian velocity at $r$ \\
$\vkz$ & Keplerian velocity at $\Rz$ \\
$\Hz$ & Gas scale height at $\Rz$ \\
$\phiz$ & Square of the aspect ratio $\Hz/\Rz$ at $\Rz$ \\
$\etaz$ & Sub-Keplerian parameter at $\Rz$ \\
$s$ & Grain size \\
$S$ & Dimensionless grain size \\
$\sz$ & Initial dimensionless grain size \\
$y$ & Grain size exponent in the drag force  \\
$\textbf{v}_{\mathrm{g}}$ & Gas velocity \\
$\textbf{v}$ & Grain velocity \\
$\rhod$ & Dust intrinsic density \\
$\md$ & Mass of a dust grain \\
$\ts$ & Drag stopping time \\
$\tsz$ & Drag stopping time at $\Rz$ \\
\hline
\end{tabular}
\end{center}
\caption{Notations used in the article.}
\label{tabnote}
\end{table}
%

\section{Settling with different growth models}
\label{app:other}

\begin{figure}
\resizebox{\hsize}{!}{\includegraphics{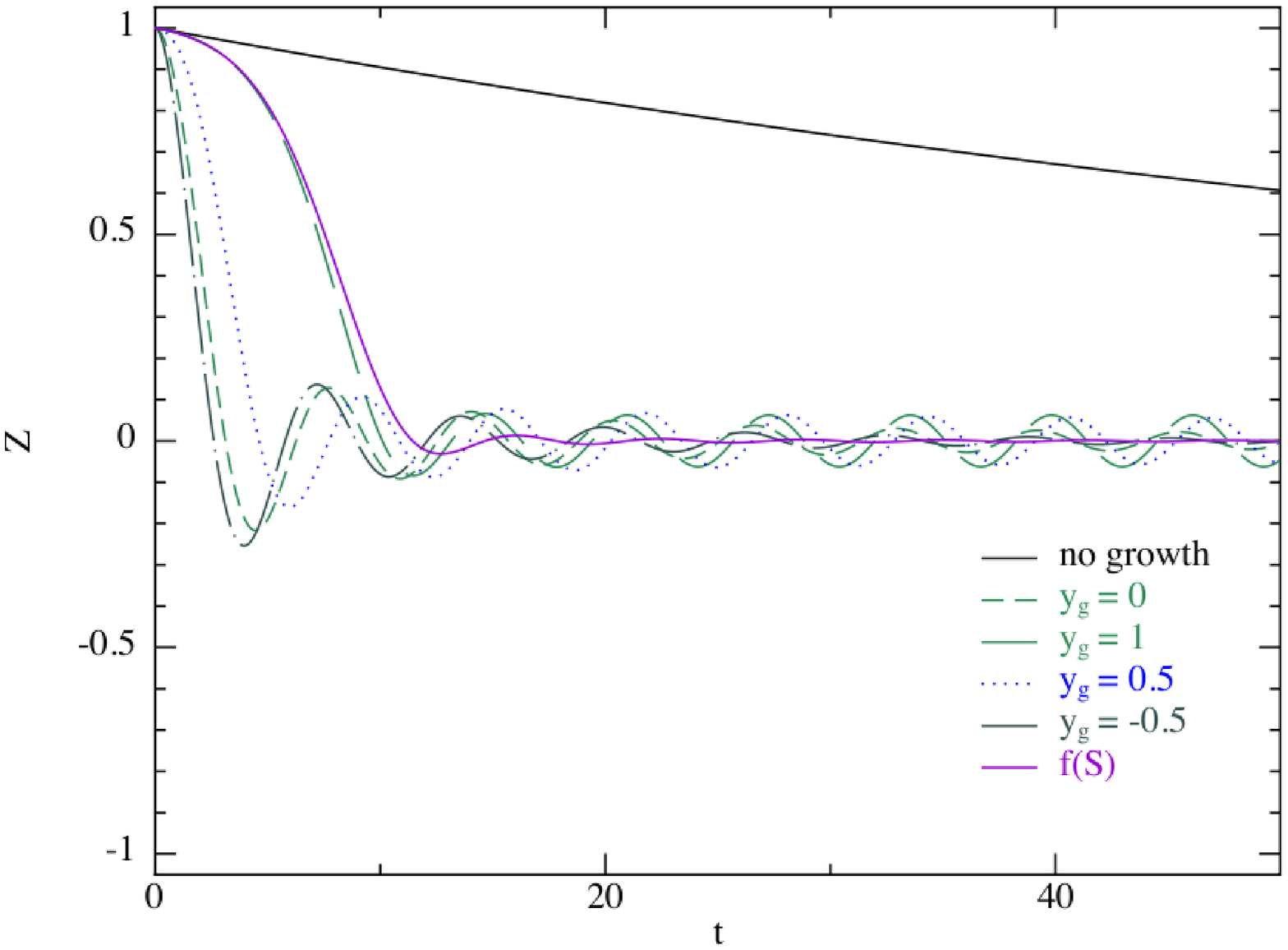}}
\caption{Vertical motion of a growing dust particle with starting at $ Z_0 = 1 $ with $ \sz = 10^{-2} $ and $ \gamma = 1/2$ for different growth models. No significant differences are found between the different models.}
\label{fig:eps5dm1}
\end{figure}

\begin{figure}
\resizebox{\hsize}{!}{\includegraphics{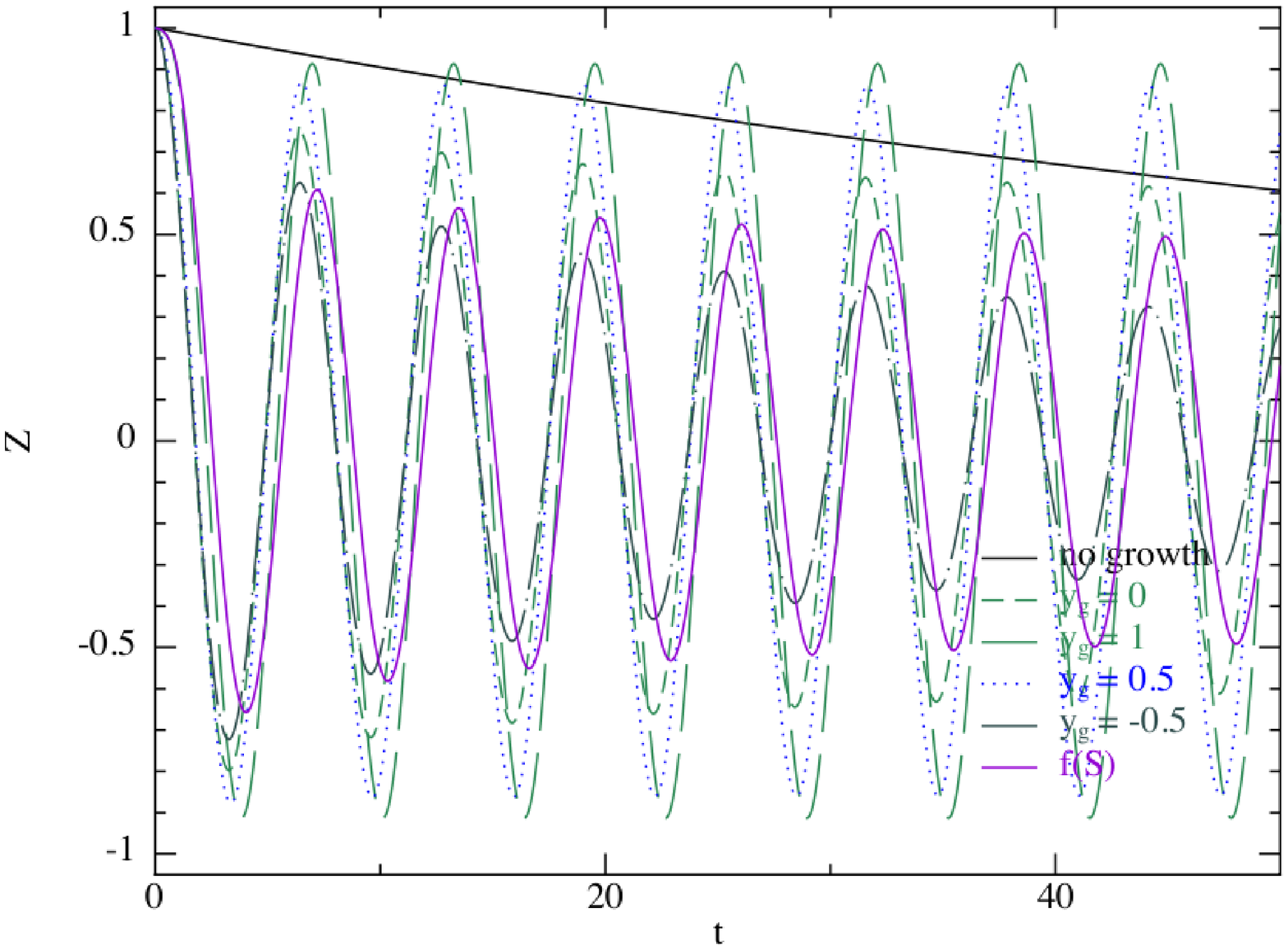}}
\caption{Vertical motion of a growing dust particle with starting at $ Z_0 = 1 $ with $ \sz = 10^{-2} $ and $ \gamma = 5$ for different growth models. No significant differences are found between the different models.}
\label{fig:eps5d0}
\end{figure}

\end{appendix}

\label{lastpage}
\end{document}